\documentstyle[preprint,aps,epsf]{revtex}
\begin{document}
\preprint{CPT-97/P.3496}
\title{Atomic parity violation in cesium and implications for new physics}
\author{Aldo Deandrea}
\address{Centre de Physique Th\'eorique\footnote{Unit\'e Propre de 
Recherche 7061}, CNRS Luminy, Case 907, F-13288 Marseille Cedex 9}

\date{May 1997}
\maketitle

\begin{abstract}
Recent high-precision measurements of atomic parity-nonconserving 
transitions between the 6S and 7S states of cesium allow for a 
determination of the weak nuclear charge with a precision of 1.3\%,
providing an improved test of the standard model at low energy. 
Implications for new physics are examined in terms of low energy effects
on the weak charge, in particular contact interactions and scalar 
leptoquark limits. Prospects for further improvements are described. 
\end{abstract}
\pacs{12.15.Ji,13.85.Rm}

\section{Introduction}
Nuclei and electrons are bound by electromagnetic interactions that
do not violate parity. However the exchange of a Z gauge boson between 
the nucleus and the atomic electrons is parity violating. The dominant 
contribution arises from the vector coupling to the nucleus with the 
axial-vector coupling to the electrons, which allows normally forbidden 
transitions.
The nuclear vector current is conserved and the nucleus acts as a source
of the weak charge $Q_W$, which is a linear combination of the Z vector
coupling to up and down quarks:
\begin{equation}
Q_W= -2 [ C_{1u} (2Z+N) + C_{1d} (Z+ 2N)]
\end{equation}
where $N$ is the number of neutrons, $Z$ the number of protons and
\begin{equation}
C_{1u}= -\frac{1}{2} + \frac{4}{3} \sin^2 \theta_W, \;\;\;
C_{1d} =\frac{1}{2} - \frac{2}{3} \sin^2 \theta_W,
\end{equation}
at tree-level. 
The tree-level formula for the weak charge is modified by 
radiative corrections \cite{radc} and by physics  beyond the
standard model, thus a precise determination of $Q_W$ provides a test
of these corrections and bounds on new physics, complementing the
results of high energy colliders. Theoretically the structure of cesium is 
the most accurately known ($1\%$) \cite{atom} among heavy atoms, as  
it is an alkali-metal atom and it can be described as a valence electron 
and a closed-shell tightly bound core which is relatively unpolarizable.

Recently a factor of 7 improvement in the measurement of parity-nonconserving
transitions between the 6S and 7S states of $^{133}Cs$ was reported
\cite{wood} with the use of a spin polarized atomic beam. The nucleus has 
spin $I=7/2$ and the total angular momentum of the atomic $S$ states is 
then 3 or 4. The experiment measures both $6S(F=3) \to 7S(F=4)$ and 
$6S(F=4) \to 7S(F=3)$ transitions. The two measured transitions differ because
of nuclear spin-dependent effects. A linear combination of the two values
eliminates nuclear spin-dependent contributions, allowing a precise
determination of the weak charge. The result of \cite{wood} is:
\begin{equation}
Q_W= -72.11 \pm 0.27 \pm 0.89 
\end{equation}
where the first error is experimental and the second one is the consequence
of atomic theory uncertainty. Contrary to previous data \cite{noe}, the
theoretical uncertainty dominates the error. 

The other linear combination allows a determination of the spin-dependent
effects, mainly due to the anapole moment \cite{zel}. However it involves 
strong interaction uncertainties and I will not discuss it in the following.

Assuming the validity of the standard model one can extract the Weinberg
angle from $Q_W$, in order to have an idea of the constraints implied
by the new cesium data. Using the formulae of \cite{radc,blan} and the 
updated analysis of \cite{jeg} concerning the hadronic vacuum polarization
corrections to $\gamma Z$ mixing, in the $\overline {MS}$ scheme at the 
scale $m_Z$, and with a Higgs mass of 300 GeV, I find 
$\sin^2 \theta_W (m_Z)_{\overline{MS}} = 0.2267 \pm 0.0040$.
In order to compare this result with LEP data one should calculate
instead $\sin^2 \theta_{eff}$, which is the usually quoted value, but 
the difference is tiny \cite{gamsir}, and one can safely neglect it
in the following considerations. From the previous determination
one can see that in $\sin^2 \theta_W (m_Z)_{\overline{MS}}$
the error is reduced by a factor of two with respect to previous 
determinations from cesium \cite{pdg}. 
Now the error is dominated by the atomic physics theoretical
accuracy \cite{atom,blun}, which may improve in the near future to 
few parts in $10^3$. The present error in the $\sin^2 \theta_W$ 
determination is an order of magnitude larger than the one quoted by 
LEP, $\sin^2 \theta_{eff}=0.23200(27)$ \cite{alt}, but cesium is likely
to play a more significant role in tests of the standard model in 
the future. For example assuming the atomic theory error to go down
to 0.2\% and the experimental error to stay the same, the induced error
in the determination of $\sin^2 \theta_W$ would be 0.0013.

\section{Contact interactions}

Recently much interest was devoted to contact interactions and scalar
leptoquarks as they might account for the excess of high $Q^2$ events 
at HERA \cite{hera} in $e^+p$ collisions. 
Atomic parity violation puts severe bounds on quark-lepton four-fermion 
contact interactions \cite{lan,desh}. The relevant lagrangian is written
in the form \cite{pes}:
\begin{equation}
{\cal L}_{4f}=\sum_{i,j=L,R; q=u,d} \frac{4\pi \epsilon^q_{ij}}
{(\Lambda_{ij}^q)^2} {\bar e}_i \gamma_{\mu} e_i 
{\bar q}_j \gamma^{\mu} q_j
\end{equation}
where $\epsilon_{ij}=\pm 1$. The contact interaction produces a shift 
of the weak charge:
\begin{eqnarray}
\delta Q_W &=& - (143 {\mathrm TeV}^2) \left(
\frac{\epsilon^u_{RL}}{{\Lambda^u_{RL}}^2}
-\frac{\epsilon^u_{LL}}{{\Lambda^u_{LL}}^2}
-\frac{\epsilon^u_{LR}}{{\Lambda^u_{LR}}^2}
+\frac{\epsilon^u_{RR}}{{\Lambda^u_{RR}}^2} \right) \nonumber \\
&-& (160 {\mathrm TeV}^2) \left(
\frac{\epsilon^d_{RL}}{{\Lambda^d_{RL}}^2}
-\frac{\epsilon^d_{LL}}{{\Lambda^d_{LL}}^2}
-\frac{\epsilon^d_{LR}}{{\Lambda^d_{LR}}^2}
+\frac{\epsilon^d_{RR}}{{\Lambda^d_{RR}}^2} \right)
\label{dql}
\end{eqnarray}
and using the experimental result 
for the weak charge \cite{wood} and the theoretical standard model
value  $-72.88 \pm 0.06$ of \cite{pdg}, eq. (\ref{dql})
gives the limits of table I, which are larger 
than those coming from high-energy colliders (see for
example \cite{pierre} and recent bounds by CDF \cite{cdf} and LEP 2 \cite{lep2}). One should be careful, when comparing table I
with the bounds reported in the literature, that the definition of
$\Lambda^q_{ij}$ may differ by a factor of $\sqrt{4\pi}$.
The limit coming from eq. (\ref{dql}) can however 
be eluded if the contact interactions are parity conserving or 
cancellations occur (see the discussion in \cite{ann}). 

\section{Scalar leptoquarks}

Let us now consider the low energy effects of scalar leptoquarks that couple
chirally and diagonally to the first generation. ``Chiral'' means that
the leptoquark is coupled either to left-handed or to right-handed quarks
but not to both, in order to avoid unacceptable deviations from lepton
universality, for example in $\pi \to e\nu$, while ``diagonal'' 
means that the leptoquark 
couples only to a single leptonic and quark generation (at least
approximately as the CKM matrix induces mixing for left-coupled
leptoquarks, and as a consequence flavour changing neutral currents).
 
I follow here the notation of \cite{mir}, where previous cesium 
data was used to set bounds on the leptoquarks. From the general effective
lagrangian satisfying baryon and lepton number conservation, with the
most general dimensionless and $SU(3)_c \times SU(2)_L \times U(1)_Y$ 
invariant couplings \cite{buc} one has scalar leptoquarks with the 
following quantum numbers with respect to $SU(3)_c \times SU(2)_L 
\times U(1)_Y$: $S(3^*,1,1/3)$, ${\tilde S} (3^*,1,4/3)$, $T(3^*,3,1/3)$,
$D(3,2,7/6)$, ${\tilde D}(3,2,1/6)$. $\tilde S$ and $\tilde D$ couple 
only to right-handed quarks, $T$ only to left-handed quarks, while $S$
and $D$ can couple to both. 

In a class of supersymmetric theories it is natural to have $S_L$, 
$S_R$ of type $S$ and $D_L$, $D_R$ of type $D$ \cite{slep}, 
where the subscripts 
$R$ and $L$ refer to the helicity of the quark coupled to the 
leptoquark. The requirement of having chiral couplings 
in the sense indicated before may therefore be fulfilled, if mixing
between the left- and right-coupled leptoquarks is small. 

In the following I shall be mainly interested in $D_L$, $D_R$ and 
$\tilde D$ leptoquarks, which are relevant in the discussion of 
the excess of high $Q^2$ events at HERA in $e^+p$ collisions.
Their Yukawa interactions are given by
\begin{eqnarray}
{\cal L} &=& g_R ({\bar e} u_R D_R^{(-5/3)} + {\bar \nu} u_R D_R^{(-2/3)})
+ g_{\tilde D} ({\bar e} d_R {\tilde D}^{(-2/3)} + {\bar \nu} d_R
{\tilde D}^{(1/3)}) \nonumber \\ 
&+& g_L ({\bar e}u_L D_L^{(-5/3)} + 
{\bar e}d_L D_L^{(-2/3)}) + c.c.
\end{eqnarray}
where for the $D_L$ I neglected second and third generation
couplings, as the leptoquark is assumed to be diagonal to a good
approximation. The superscript in parentheses in the previous 
formula refers to the leptoquark electromagnetic charge.

One can easily derive the effective low energy 4-fermion interaction 
due to leptoquark exchange and obtain the corresponding shift in the 
weak charge. In order to have a compact notation, it is useful to 
introduce the parameter $\eta_I$ which has the value 1 when the
leptoquark $I$ is present and otherwise is zero. The contribution
to $Q_W$ is:
\begin{eqnarray}
\delta Q_W^{lq} &=& -\frac{1}{2 {\sqrt 2} G_F} \left( \frac{g}{M}
\right)^2 [(-\eta_{S_L} +\eta_{S_R} -\eta_{D_L} + \eta_{D_R} -\eta_T)
(2Z+N) \nonumber \\ 
&+& (\eta_{\tilde S} -\eta_{D_L} + \eta_{\tilde D} -2\eta_T)
(Z+2N)]
\label{qwlq}
\end{eqnarray}
where $G_F$ is the Fermi constant. Using the experimental result 
for the weak charge \cite{wood} and the theoretical standard model
value  $-72.88 \pm 0.06$ of \cite{pdg}, one obtains the limits of
table II, in terms of the ratio leptoquark mass $M$ over coupling $g$.

Limits from other experimental data can further constrain the 
leptoquarks, depending on the nature of the leptoquark, for example
flavour changing neutral current limits or universality in leptonic
$\pi$ decay. Moreover note that the limits coming from 
eq.(\ref{qwlq}) are valid when one assumes that there is only one
leptoquark multiplet and that there is no mass splitting within
the multiplet (for a detailed discussion see appendix B of \cite{mir}).
The presence of more than one leptoquark multiplet can both improve
or reduce the bound, as different leptoquarks may contribute
to the weak charge with the same or opposite sign.

I will not analyse the implications of leptoquarks for the high-$Q^2$
HERA data in detail, as this is the subject of dedicated papers \cite{al2}.
Note however that $D_{L,R}$ and $\tilde D$ leptoquarks are produced in
$e^+p$ collisions in the s-channel, whereas they contribute to
$e^-p$ non-resonantly. If a mass of 200 GeV is assumed for the 
leptoquarks, a bound on the couplings can be derived, 
\begin{equation}
g_{L,R} < 0.09 ~~~~~~~~~~~g_{\tilde D} <0.08
\end{equation} 
at 95 \% C.L. while the coupling required to 
explain the HERA anomaly is lower, of the order of 0.02 for a scalar
leptoquark coupled to $u$ quarks (like $D_R$), and 0.04 if coupled 
to $d$ quarks (like $\tilde D$) and thus within the permitted region.

Limits form cesium can be combined with direct leptoquark searches
to perform a two parameter fit in the mass and coupling
of the leptoquarks. In the following $\chi^2$ analyses a small number
of events has to be compared with theory. I use for them the method of
least squares with a $\chi^2$ function for Poisson-distributed data,
which asymptotically behaves like a classical $\chi^2$ \cite{stat}. The 
goodness-of-fit (confidence level) is evaluated approximatively 
as if data were Gaussian-distributed.

Leptoquark production at Tevatron is insensitive
to the $g$ coupling (the only requirement is $g > 10^{-12}$ due to the
event reconstruction algorithms used in the experimental analysis), 
but detection of the leptoquark decay products depends on the 
leptoquark branching fraction to $\ell q$ and $\nu q$. A recent 
analysis from the D0 collaboration was used in the fit \cite{d0}. 
The leading order parton production cross-section \cite{blu} was 
convoluted with parton distribution functions \cite{mrsa}, at a 
scale $\mu=M$ (this choice of the scale at leading order is supported
by the next-to-leading result of \cite{kra}), in order to produce the theoretical cross section at Tevatron.
The result of the fit is shown in Fig.1, as a 95\% C.L. bound on the mass
and coupling of the $\tilde D$ scalar leptoquark (limits for $D_L$ and $D_R$
are similar and are not shown). It is assumed that $\tilde D$ has
100\% branching fraction to $e q$. Fig. 2 shows the same limits assuming
50\% branching to $e q$ and 50\% to $\nu_e q$. 

Assuming that the excess of events seen at high-$Q^2$ in $e^+p$
collisions at HERA \cite{hera} is due to the production of a scalar 
leptoquark, I calculated bounds in the plane ($M$,$g$) using cesium data 
\cite{wood} and the leading order parton cross-section for scalar
leptoquark production \cite{buc} with parton 
densities \cite{mrsa}, taking into account initial state photon 
radiation from the positron. The combined integrated luminosity 
of 34.3 pb$^{-1}$ from H1
and ZEUS collaborations at ${\sqrt s}=300$ GeV in the $e^+p \to eX$
mode was used, taking into account that the two experiments have seen
24 events with $Q^2 > 15000$ GeV$^2$, while the standard model expectation
is $13.4 \pm 1$ events. Fig.3 refers to a $\tilde D$ leptoquark, 
while Fig.4 to a $D_R$ leptoquark. The limits for a $D_L$ leptoquark
assuming that the members of the doublet are degenerate in mass, are
similar to those of Fig.4 and are not shown. No information was
given to the fit on the invariant mass distribution of the anomalous HERA
events, in order to see if the preferred ($M$,$g$) values were consistent
with those required by the leptoquark interpretation of HERA anomalous
events. They turn out to be consistent, however $\chi^2$ is small in 
a narrow band over a wide range of masses. Note that if the contour plot is performed using a too coarse-grained grid of points, this can be missed.
One can at most say that for the $\tilde D$ leptoquark the preferred
region is for $M \lesssim 220$ GeV, $g \lesssim 0.06$ (for a mass of 
200 GeV the preferred value of $g$ is 0.034) and for the $D_R$ leptoquark 
$M \lesssim 250$ GeV, $g \lesssim 0.07$ (for a mass of 200 GeV the 
preferred value of $g$ is 0.017). The preferred region for the fit is
obtained demanding $\chi^2 < 2$, note however that for low $\chi^2$
the Gaussian confidence-level estimate is not appropriate.

\noindent
\vspace*{0.2cm}
\noindent
{\bf Acknowledgments}
\par\noindent
The author acknowledges the support of a TMR research fellowship of the 
European Commission under contract nr. ERB4001GT955869, the kind 
hospitality of the Dipartimento di Fisica, Universit\`a di Pavia, where
part of this work was done, and a discussion on Tevatron limits with
P.Chiappetta. Numerical calculations were performed using Mathematica 
\cite{math}.

\begin{table}
\hfil
\vbox{\offinterlineskip
\halign{&#& \strut\quad#\hfil\quad\cr
\tableline
\tableline
&\omit&& $\epsilon^u=1$ && $\epsilon^u=-1$ && $\epsilon^d=1$ && 
$\epsilon^d=-1$ & \cr
\tableline
&\omit&&\omit&&\omit&&\omit&&\omit&\cr
& LL, LR && 7.4 && 11.7 && 7.9 && 12.3 & \cr
& RR, RL && 11.7 && 7.4 && 12.3 && 7.9 & \cr
&\omit&&\omit&&\omit&&\omit&&\omit&\cr
\tableline
\tableline}}
\vspace*{1cm}
\par\noindent
\caption{95\% C.L. lower bounds from contact interactions on 
$\Lambda^q_{ij}$ in TeV units
assuming only one of the $\epsilon^q_{ij}$ terms is responsible for
the shift in the weak charge.}
\end{table}

\vspace*{2cm}
\begin{table}
\hfil
\vbox{\offinterlineskip
\halign{&#& \strut\quad#\hfil\quad\cr
\tableline
\tableline
&\omit&& $S_L$ && $S_R$ && $D_L$ && $D_R$ && $T$ && $\tilde S$ && 
$\tilde D$ & \cr
\tableline
&\omit&&\omit&&\omit&&\omit&&\omit&&\omit&&\omit&&\omit&\cr
&$M/g$&&1500&&2300&&2200&&2300&&2700&&2500&&2500& \cr
&\omit&&\omit&&\omit&&\omit&&\omit&&\omit&&\omit&&\omit&\cr
\tableline
\tableline}}
\vspace*{1cm}
\par\noindent
\caption{95\% C.L. lower bounds on scalar leptoquarks in GeV units
for the ratio $M/g$.}
\end{table}

\begin{figure} 
\epsfxsize=7truecm
\centerline{\epsffile{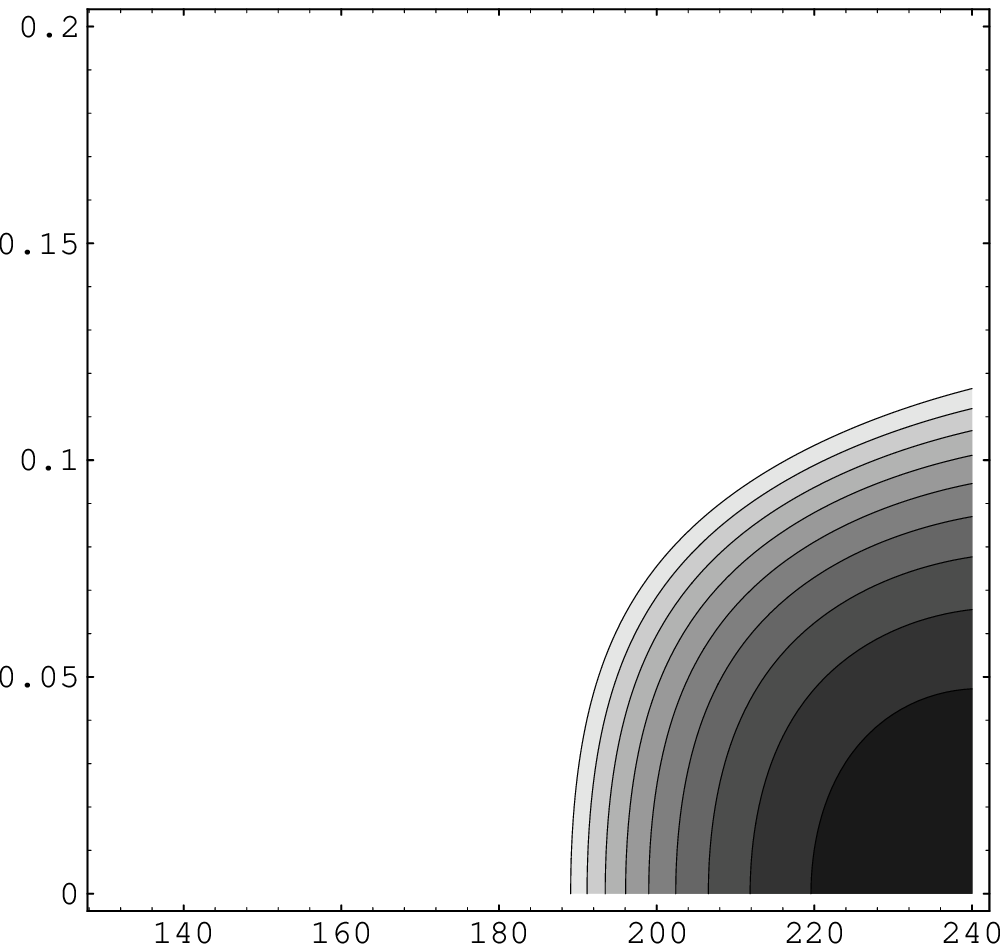}}
\noindent
{\bf Fig. 1} - {$95 \%$ C.L. limit in the mass $M$ (GeV) and coupling $g$ 
of a $\tilde D$ leptoquark from atomic parity violation in cesium
\protect{\cite{wood}} and the direct search at TEVATRON (data from the D0
collaboration \protect{\cite{d0}}), assuming a branching ratio of the
leptoquark to $ed$ equal to 100\%. The allowed region in the parameter
space is the grey area. Darker regions correspond to lower $\chi^2$
values.}
\end{figure}

\begin{figure} 
\epsfxsize=7truecm
\centerline{\epsffile{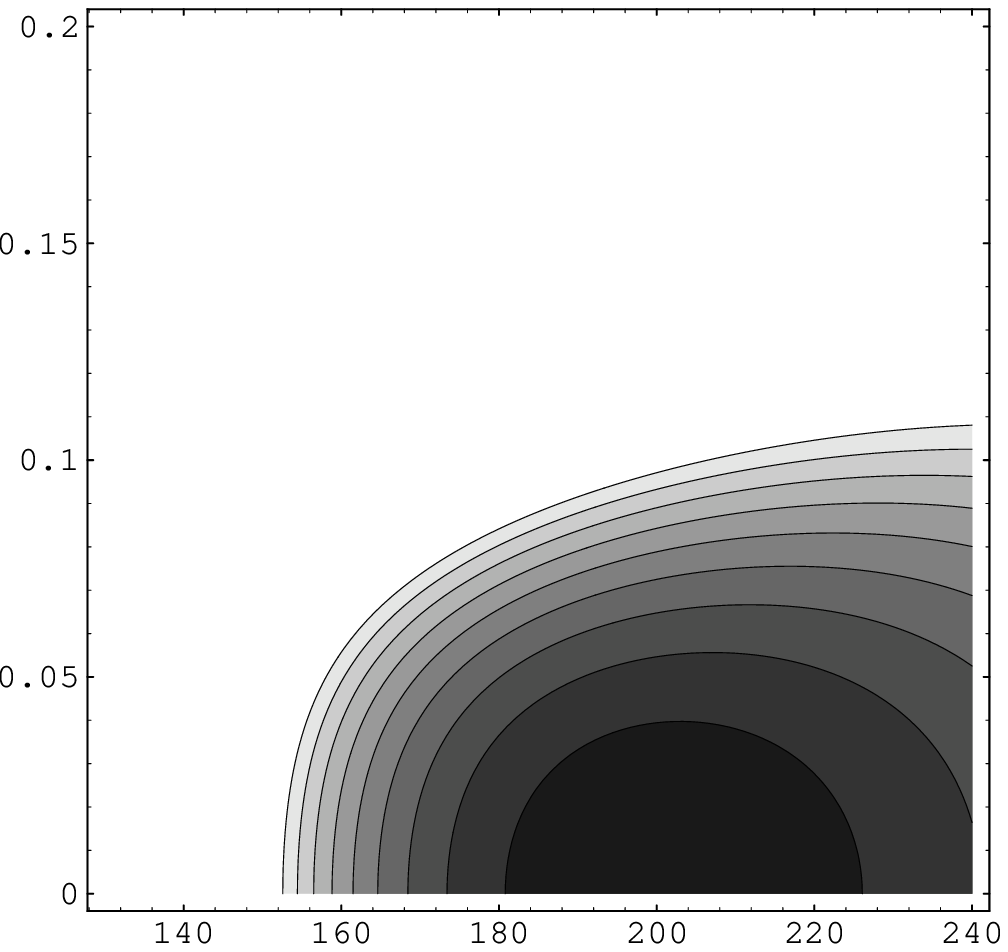}}
\noindent
{\bf Fig. 2} - {$95 \%$ C.L. limit in the mass $M$ (GeV) and coupling $g$ 
of a $\tilde D$ leptoquark assuming a branching ratio of the
leptoquark to $ed$ of 50\% and to $\nu_e d$ of 50\%.}
\end{figure}

\begin{figure} 
\epsfxsize=7truecm
\centerline{\epsffile{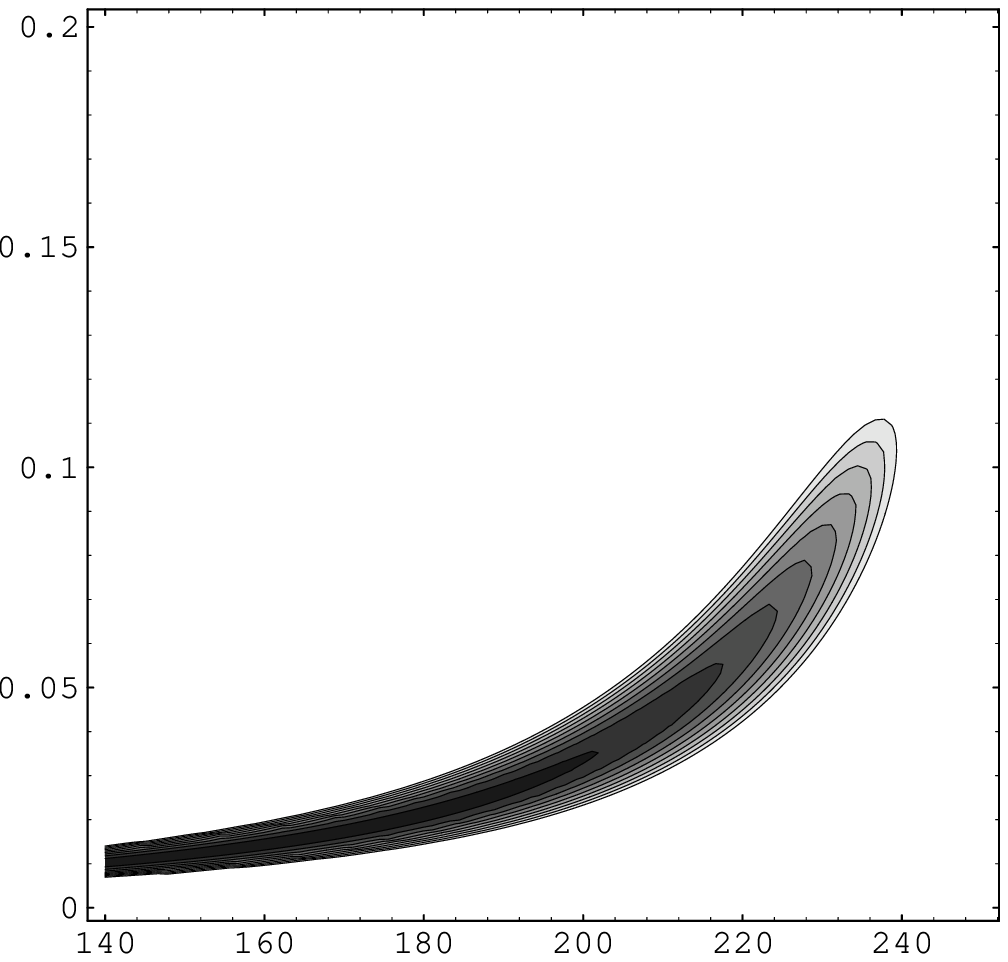}}
\noindent
{\bf Fig. 3} - {$95 \%$ C.L. limit in the mass $M$ (GeV) and coupling $g$ 
of a $\tilde D$ leptoquark from the combination of cesium data and the 
cross-section excess at HERA.}
\end{figure}

\begin{figure} 
\epsfxsize=7truecm
\centerline{\epsffile{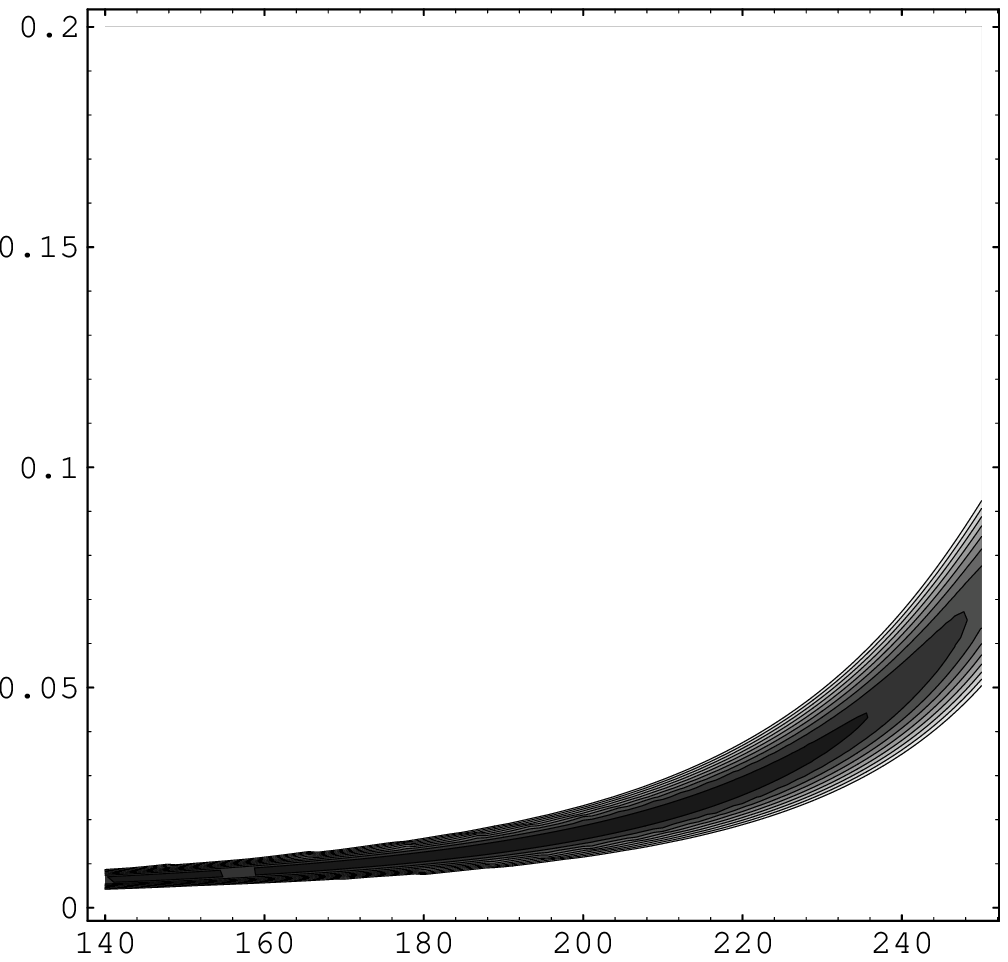}}
\noindent
{\bf Fig. 4} - {$95 \%$ C.L. limit in the mass $M$ (GeV) and coupling $g$ 
of a $D_R$ leptoquark from the combination of 
cesium data and the cross-section excess at HERA.}
\end{figure}

\end{document}